\documentclass[twocolumn,showpacs,preprintnumbers,amsmath,amssymb,superscriptaddress]{revtex4}
\usepackage{amsfonts,amssymb,bm,graphicx,times} \usepackage{amsmath}
\usepackage{float,color}

\begin{document}

\title{Perfect excitation of a matter qubit by a single photon in free
  space}

\author{Magdalena Stobi\'nska}
\email{mstobinska@optik.uni-erlangen.de} \affiliation{Institut f\"ur
  Optik, Information und Photonik, Max Planck Forschungsgruppe,
  Universit\"at Erlangen-N\"urnberg, G\"unter-Scharowsky-Str.\ 1, Bau
  24, 91058 Erlangen, Germany}

\author{Gernot Alber} \affiliation{Institut f\"ur Angewandte Physik,
  Technische Universit\"at Darmstadt, 64289 Darmstadt, Germany}

\author{Gerd Leuchs} \affiliation{Institut f\"ur Optik, Information
  und Photonik, Max Planck Forschungsgruppe, Universit\"at
  Erlangen-N\"urnberg, G\"unter-Scharowsky-Str.\ 1, Bau 24, 91058
  Erlangen, Germany}

\begin{abstract}
We propose a scheme for perfect excitation of a single two-level atom
by a single photon in free space.  The photon state has to match the
time reversed photon state originating from spontaneous decay of a
two-level system. We discuss its experimental preparation.  The state
is characterized by a particular asymmetric exponentially-shaped
temporal profile. Any deviations from this ideal state limit the
maximum absorption.  Although perfect excitation requires an infinite
amount of time we demonstrate that there is a class of initial
one-photon quantum states which can achieve almost perfect absorption
even for a finite interaction time.  Our results pave the way for
realizing perfect coupling between flying and stationary qubits in
free space thus opening a possibility for building scalable quantum
networks.
\end{abstract}

\maketitle

Efficient coupling between light and matter at a single quantum level
lies at the heart of scalable quantum information processing,
computation and communication
\cite{Cirac1997,Duan2001,Maitre1997}. Information encoded in a flying
qubit used for its transfer has to be recorded by a localized
stationary qubit (e.g. an atom), i.e. the photon has to excite the
atom with unit probability.  Experimental realization of these
protocols remains challenging due to the weak coupling between a
single-photon and a single-atom in free space.  Recent
approaches investigating this problem focus on 
the absorption of a single photon
by an ensemble of atoms resulting in a distributed single photon
excitation entangling the atoms of the
ensemble~\cite{Kuzmich2003,Wal2003,Balic2005,Scully2006,Molmer2008,Glauber}.

So far, close-to-perfect interaction has been achieved only in the
context of cavity QED in the strong coupling regime where the atom is
forced to interact with a single mode of the radiation field only
\cite{Walther,Haroche,Kimble,Rempe,Savage,Fortunato,CiracSoto}.
Scaling up these schemes is difficult because of the requirement for
high finesse cavities.

Currently, several groups are attempting to quantify
\cite{Gerhardt,Sandoghdar} and to improve light-matter coupling in the
absence of any mode-selecting cavity \cite{Tey,Sondermann2007}.  A
detailed study of single-atom-single-photon interaction in free space
requires the control of all resonant degrees of freedom of the
radiation field, i.e. its spatio-temporal vector modes.  Van Enk and
Kimble showed theoretically that both strong focusing and increased
overlap of a light beam with a dipole wave corresponding to the
relevant atomic transition improve the coupling
\cite{Enk2001,Enk2004}. Strong focusing of a light beam was
demonstrated by Quabis et al. by tailoring the polarization pattern of
light in theory \cite{Quabis2000} and in experiment
\cite{Dorn2003,Quabis2005}.

We aim at having full control over the field modes and at exciting and
maximizing the coupling to the atom.  A first step towards this goal
was the demonstration of a significant attenuation of a laser beam by
a single trapped ion \cite{Wineland}. Recently several groups
succeeded in improving on this result
\cite{Gerhardt,Vamivakas,Tey,Zumofen,Domokos}.  Other groups attempted
to control the excitation of a single atom \cite{Weinfurter,Grangier},
the ultimate goal being perfect excitation with a single photon wave
packet \cite{Enk2004,Sondermann2007, Pinotsi}. This should be possible
based on a time reversal argument applied to spontaneous emission of a
single photon \cite{Quabis2000}. These two goals however, namely
maximum attenuation of light and prefect excitation of the atom, are
distinctly different. On the one hand, attenuation of a weak laser
beam with zero transmission should be reachable with the atom
populating essentially the ground state and with narrow-band
continuous-wave laser radiation. On the other hand, a perfectly
excited atom cannot be realized in dynamical equilibrium but only in a
transient process which involves the excitation by a $\pi$--pulse, for
example.  Our objective is to realize the equivalent of a $\pi$--pulse
excitation process with a single photon wavepacket \cite{Raymer}
properly shaped in space and time.

In this Letter we outline a scheme for perfect excitation of a single
two-level atom by a single photon in free space.  We show that for
this purpose the photon state has to match the time reversed photon
state originating from spontaneous decay of a two-level system and we
discuss its experimental preparation.  Any deviations from this ideal
state limit the maximum absorption.  Both the spontaneous decay
process and its reverse process of perfect excitation require an
infinite amount of time.  However, we demonstrate that there is a
class of one-photon quantum states with mode decompositions close to
the ideal one which can achieve almost perfect absorption even for a
finite interaction time.  As the mean values of the electric and the
magnetic field of the radiation field are zero, the excitation of the
matter qubit is completely caused by its uncertainties.  They reveal
an exponentially increasing tail and a sharp-edged shape genuine to a
time reversed dipole wave \cite{Quabis2000,Enk2004}. In contrast to
other recent work, such as Ref.\cite{Zumofen,Tey} which is based on
classical diffraction theory, we apply a full non-relativistic QED
treatment.

\textit{Theory:} Let us consider a trapped atomic qubit at a fixed
position ${\bf x}\!=\!0$ in free space.  Its excited and ground state are
denoted by $|e\rangle$ and $|g\rangle$, respectively.  Temporally, the
atom interacts almost resonantly with the quantized radiation field
which contains only one photon distributed over a continuum of
temporal modes centered around the optical atomic transition frequency
$\omega_0$.  Its center-of-mass motion is not affected if the atom is
cooled to its lowest vibrational state so that its recoil momentum is
picked up by the atom-trap system as a whole, much like in the
M\"ossbauer effect. In the interaction picture the Hamiltonian of this
matter-field system reads
\begin{equation}
  \hat{V}(t) = \hbar \sum_{l\in I} g_l \hat{\sigma}_+ \hat{a}_l e^{i
    \Delta_l(t-t_0)} + \mathrm{h. c. }
\label{Ham}
\end{equation}
where $\Delta_l \!=\! \omega_0 - \omega_l$ is the detuning between the
atom and the $l$-th optical mode with frequency $\omega_l$ and $g_l\!
=\!  -({i}/{\hbar})\sqrt{({\hbar\omega_l})/({2\epsilon_0})} \, {\bf
  d}\cdot {\bf u}_l({\bf x \!=\!0})$ is the $l$-th mode coupling
constant.  The orthonormal mode functions of the radiation field are
denoted by ${\bf u}_l({\bf x})$ and ${\bf d}$ is the atomic dipole
moment.  The annihilation and creation operators of the field modes
are given by $\hat{a}_l$ and $\hat{a}_l^{\dagger}$.  Similarly,
$\hat{\sigma}_+ \!\!=\!\! |e\rangle \langle g|$ and $\hat{\sigma}_-
\!\!=\!\!  |g\rangle \langle e|$ are the atomic rising and lowering
operators. These latter operators are expressed in the Schr\"odinger
picture which coincides with the interaction picture at time~$t_0$.
Only those radiation field modes $l\!\!\in \!\!I$ contribute
significantly to the system dynamics which are almost resonantly
coupled to the atom within a frequency interval of the order of its
spontaneous decay rate $\Gamma \!= \!(4/3)\omega_0^3
|\mathbf{d}|^2/(4\pi \epsilon_0 \hbar c^3) \! \ll \! \omega_0$.

We aim at determining the particular one-photon state of the quantized
radiation field in free space which enables perfect excitation of the
two-level atom prepared in its ground state $|g\rangle$ initially.  In
the interaction picture the normalized quantum state of the coupled
atom-field system is of the form
\begin{eqnarray}
|\psi(t)\rangle &=& \sum_{l \in I} f_l(t) |g\rangle\otimes
\hat{a}_l^{\dagger}|0\rangle + f_0(t) |e\rangle \otimes |0\rangle
\label{state}
\end{eqnarray}
with $|0\rangle $ denoting the vacuum of the radiation field and with
the normalization constraint $|f_0(t)|^2 + \sum_{l \in I} |f_l(t)|^2
=1$.  The ideal one-photon wave packet is determined by a solution of
the time dependent Schr\"odinger equation
\begin{equation}
i \hbar \, \partial_t |\psi(t)\rangle = \hat{V}(t) |\psi(t)\rangle
\label{Schrodinger}
\end{equation}
in the time interval $(-\infty,t_0]$ with initial condition
  $f_0(t_0)\!\!=\!\!1$, $f_l(t_0)\!\!=\!\!0$.  Within the
  Wigner-Weisskopf approximation~\cite{Scully} the unique solution of
  Eq.~(\ref{Schrodinger}) is given by
\begin{eqnarray}
\label{amplitudes}
f_0(t) \!\!&=& \!\! e^{-\Gamma |t-t_0|/2},\\ f_l(t) \!\! &=& \!\!
\frac{g_l^*}{\Delta_l - i \frac{\Gamma}{2} {\rm sgn(t-t_0)}}
\left(e^{-{\Gamma}|t-t_0|/{2}}e^{-i\Delta_l(t-t_0)}-1\right),\nonumber
\end{eqnarray}
with ${\rm sgn}(x) \!\!= \!\!x/|x|$ for $x\!\!\neq \!\!0$. This
solution is valid for time $t\in(-\infty,t_0]$ or $t\in [t_0,\infty)$
    and it is continuous but not smooth at time $t_0$ where the
    specified conditions are met.  According to Eqs.~(\ref{state}) and
    (\ref{amplitudes}) the atom-field state is separable only at times
    $t \!\!\to \!\!\pm \infty$ with the atom being in its ground
    state.  Thus, in the time interval $(-\infty,t_0]$
  Eq.~(\ref{amplitudes}) describes perfect absorption of the photon
  which is completed at time $t_0$.  Similarly, in the interval
  $[t_0,\infty)$ it describes spontaneous decay of the two-level
    system initially prepared in its excited state at $t_0$.

From the uniqueness of the solution (\ref{amplitudes}) it follows that
perfect excitation of the two-level atom starting from its ground
state is possible only if the light is prepared in the one-photon
state (expressed in the Schr\"odinger picture)
\begin{eqnarray}
|\chi\rangle_- = -\lim_{t_{in}\to-\infty}\sum_{l\in I}
\frac{g_l^*}{\Delta_l + i \frac{\Gamma}{2}}\, e^{-i\omega_l
  (t_{in}-t_0)} a_l^{\dagger}|0\rangle.
\label{initialideal}
\end{eqnarray}
The solution also shows that completion of both perfect absorption and
spontaneous emission of a photon require an infinite amount of time.
This fact is not surprising in view of the well known exponential
decay law since perfect excitation might be considered as the reverse
process of spontaneous emission.  The final state of the radiation
field resulting from spontaneous decay of the atomic qubit starting at
time $t_0$ reads
\begin{eqnarray}
|\chi\rangle_+ = -\lim_{t_{in}\to \infty}\sum_{l\in I}
\frac{g_l^*}{\Delta_l - i \frac{\Gamma}{2}}\, e^{-i\omega_l
  (t_{in}-t_0)} a_l^{\dagger}|0\rangle.
\end{eqnarray}

\begin{figure}
\begin{center}
  \scalebox{0.7}{\includegraphics{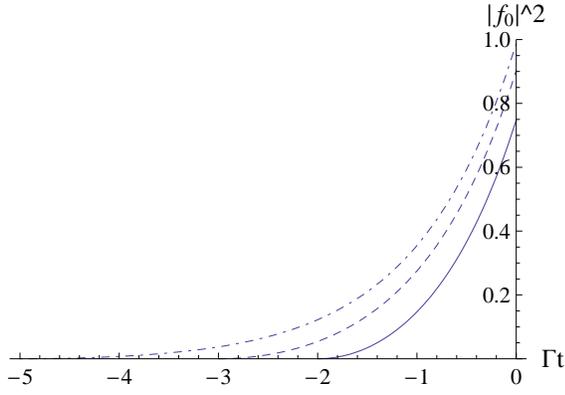}}\\
\end{center}
\caption{Excitation probability $|f_0(t)|^2$ of Eq.(\ref{f01})
  evaluated for $\Gamma T = 2$ (full), $\Gamma T = 3$ (dashed) and
  $\Gamma T = 5$ (dash - dotted): The corresponding maximum excitation
  probabilities are $0.75$, $0.90$, and $0.99$.}
\label{Fig1}
\end{figure}

In view of Eq.(\ref{initialideal}) the natural question arises what
excitation probabilities can be obtained during a finite interaction
time $t\! \in \!\![t_{in},\, t_0]$ and how do they depend on the
spatio-temporal properties of exciting wave packet?  For this purpose
let us assume that initially at time $t_{in}$ we prepare the atom in
its ground state and the radiation field in the state
\begin{eqnarray}
|\chi(t_{in})\rangle &=& - \sum_{l\in I} \frac{g_l^*}{\Delta_l + i
  \frac{\Gamma}{2}} e^{-i\omega_l (t_{in}-t_0)}
a_l^{\dagger}|0\rangle.
\label{state1}
\end{eqnarray}
This state has the same probability distribution for the occupied
modes of the radiation field as the ideal state (\ref{initialideal})
but the phases of the corresponding probability amplitudes are
different.  For $t\!\geq \!t_{in}$ the solution of the Schr\"odinger
equation yields the excited-state probability amplitude
\begin{eqnarray}
f_0(t) \!\!&=&\!\! \int_{t_{in}}^{t} dt' e^{-\Gamma(t-t')/2}\Gamma e^{-\Gamma
  (t_0 - t')/2}\Theta(t_0 - t') \nonumber\\ &=&\Theta(t_0 -
t_{in})\left( \Theta(t_0 - t) e^{-\Gamma(t_0 - t)/2}(1 - e^{-\Gamma(t
  - t_{in})}) \right. \nonumber\\ &+& \left. \Theta(t - t_0)
e^{-\Gamma(t - t_0)/2}(1 - e^{-\Gamma(t_0 - t_{in})}) \right).
\label{f01}
\end{eqnarray}
The time evolution of the excitation probability $|f_0(t)|^2$ is
depicted in Fig. \ref{Fig1}.  It approaches unity only in the limit of
$t_{in} \!\!\to \!\!-\infty$.  However, for a finite interaction time
$T \!\!=\!\!t_0 - t_{in}$, with $T\!\!\gg \!\!1/\Gamma$, the
excitation probability approaches unity exponentially 
fast, i.e.
$|f_0(t_0)|^2 = (1 - {\rm exp}(-\Gamma T))^2$.
According to Eq.~(\ref{f01}) the probability amplitude $f_0(t)$
results from a constructive interference of all probability amplitudes
associated with the excitation of the atom by the uncertainties of the
radiation field at times $t'\!\in\![t_{in},t]$.  The probability
amplitude of the field-induced excitation during a time interval
$[t',t'+dt']$ is given by $ dt'\, \Gamma \, {\rm exp}\{-\Gamma (t_0 -
t')/2\}\Theta(t_0 - t') $ and the atomic probability amplitude to
remain in the excited state $|e\rangle$ during a time interval
$[t',t]$ by ${\rm exp}\{-\Gamma (t - t')/2\}$.  This equation also
shows explicitly that the field uncertainties can excite the atom only
for $t'\!\leq \!t_0$.

The form of the quantum state of Eq.~(\ref{state1}) is crucial for
achieving almost perfect excitation during a finite interaction time.
This state corresponds to a time-reversed dipole wave
\cite{Enk2004,Sondermann2007}.  If the initial state of the light was
replaced by a reflected but not time-reversed dipole wave, such as
\begin{eqnarray}
|\sigma(t_{in})\rangle &=& - \sum_{l\in I} \frac{g_l^*}{\Delta_l - i
  \frac{\Gamma}{2}}e^{-i\omega_l (t_{in}-t_0)} a_l^{\dagger}|0\rangle,
\label{state2}
\end{eqnarray}
the resulting excitation of the atomic qubit would change drastically.
This initial condition yields the excitation amplitude
\begin{eqnarray}
f_0(t) \!\!&=& \!\!\int_{t_{in}}^{t} dt' e^{-\Gamma(t-t')/2}\Gamma
  e^{-\Gamma (t' - t_0)/2}\Theta(t' - t_0) \nonumber\\ &=&
  -\Theta(t-t_{in})\left( \Theta(t_{in} - t_0) \Gamma(t -
  t_{in})e^{-\Gamma(t - t_0)/2} \right.  \nonumber\\ &+&
  \left. \Theta(t_0 - t_{in}) \Gamma(t - t_0)e^{-\Gamma(t - t_0)/2}
  \right).
\label{time2}
\end{eqnarray}
This probability amplitude $f_0(t)$ also results from a constructive
interference of all probability amplitudes associated with the
excitation of the atom at times $t'\!\!\in\!\![t_{in},t]$. The
probability amplitude of the atomic excitation in a time interval
$[t',t'+dt']$ is given by $dt' \, \Gamma \, {\rm exp}\{-\Gamma (t' -
t_0)/2\}\Theta(t' - t_0) $ and the probability amplitude of the atom
to remain in the excited state during a time interval $[t',t]$ by
${\rm exp}\{-\Gamma (t - t')/2\}$.  The field uncertainties
responsible for the excitation are nonzero only for $t'\!\geq
\!t_0$. This implies $f_0(t)\!= \!0$ for $t_{in}\!\leq \!t\!\leq
\!t_0$.
\begin{figure}
\begin{center}
  \scalebox{0.7}{\includegraphics{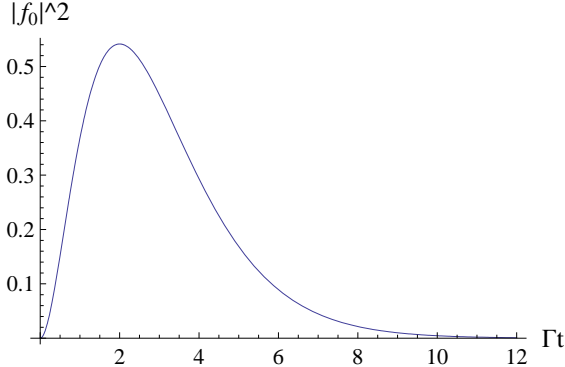}}\\
\end{center}
\caption{Excitation probability $|f_0(t)|^2$ of Eq.(\ref{time2}): The
  maximum amounts to $|f_0(t)|^2 = 0.54$.}
\label{Fig2}
\end{figure}

A typical time evolution of Eq.~(\ref{time2}) is depicted in
Fig.\ref{Fig2}. The atomic excitation reaches its maximum at the
interaction time $t_{max} \!\!= \!\!t_0 - t_{in}+2/\Gamma$ with
$t_0\geq t_{in}$ which equals $|f_0(t=t_{0} + 2/\Gamma)|^2 =
0.54$. This maximum value cannot be increased further by any choice of
the interaction time. In other words, perfect excitation cannot be
achieved with the single-photon state of Eq.~(\ref{state2}).  It is
interesting to note that the time dependence described by
Eq.~(\ref{time2}) is also obtained in the case of spontaneous decay of
an initially excited two-level atom positioned in the center of a very
large spherically symmetric metallic cavity of radius $R$ for times
$1/\Gamma \!\!\ll\!\! t\!\! \ll \!\!2t_0$~\cite{Alber1992}. Here, $t_0
\!\!= \!\!2R/c$ denotes the time which the spontaneously emitted
photon takes to return again to the atom after reflection at the
metallic boundary of the cavity. The subsequent reabsorption of the
photon is described by the probability amplitude given by
Eq.~(\ref{time2}). Recently related experimental results were reported
\cite{Blatt}.

Similarly, a single photon quantum state with a Gaussian mode
distribution
\begin{eqnarray}
|\Phi(t_{in})\rangle = \mathcal{N}\sum_{l\in I} g_l^*
e^{-\Delta_l^2/\sigma^2}e^{-i\omega_l (t_{in}-t_0)}\, a_l^{\dagger}|0\rangle,
\end{eqnarray}
will not achieve perfect excitation.  The normalization constant is
given by $\mathcal{N}\! \!= \!\!(8 \pi)^{1/4}/\sqrt{\Gamma \sigma}$
and $\sigma$ measures the width of the mode distribution.  For this
state the maximum value of the excitation probability is equal to
$|f_0(t_{max})|^2 \!\!=\!\! 0.80$ and is obtained for a width $\sigma
\!\!=\!\! 1,46\Gamma$ at time $t_{max} \!= \!t_0 + 1/\Gamma$.

It is worth noting that the interaction of a two-level atom with a
photon is governed completely by its electric and magnetic field
uncertainties since the mean values of the fields vanish for the
atom-field state (\ref{state}). The time evolution of the excitation
probability $|f_0(t)|^2$ probes the dynamics of the uncertainties of
the quantized radiation field in the immediate vicinity of the
atom. For the case of perfect absorption these uncertainties are given
explicitly by Eq.~(\ref{amplitudes}).  In order to obtain also insight
into the uncertainties of the radiation field at a space-time point
$({\bf x}, t)$ far away from the atom $| {\bf x}| \omega_0/c \!\gg
\!1$ one has to investigate the normally ordered variances of the
electric $\hat{{\bf E}}$ and the magnetic $\hat{{\bf B}}$ fields.
Before the completion of the ideal excitation process $t\!\leq \!t_0$
Eqs.~(\ref{state}) and (\ref{amplitudes}) yield the result ($\hat{{\bf
    F}} \!\!= \!\!\hat{{\bf E}}$ or $(c \hat{{\bf B}})$)
\begin{eqnarray}
\langle : \!\!\left({\bf e}\cdot\hat{{\bf F}}({\bf x})\right)^2
  \!\!\! : \rangle &=& \hbar \omega_0\, \frac{6\Gamma {\rm
      sin}^2\theta}{16\pi\epsilon_0 c|{\bf x}|^2}({\bf e}\cdot {\bf
    e}_{\theta})^2 \\ &{}& e^{-\Gamma(t_0 - t - |{\bf x}|/c)} \Theta(t_0
  - t - |{\bf x}|/c), \nonumber
\label{fluctuations}
\end{eqnarray}
in the limit of interest $\omega_0 \!\ll \!\Gamma$.  Here, $\theta$
denotes the angle between the atomic dipole operator ${\bf d}$ and
${\bf x}$, ${\bf e}_{\theta}$ and ${\bf e}_{\varphi}$ are unit vectors
along the corresponding coordinate lines of spherical coordinates and
${\bf e}$ is an arbitrary unit vector.  These variances reveal the
characteristic shape of the time-reversed dipole wave which leads to
asymptotically perfect excitation under the model assumptions made
above.

\textit{Planned experiment:} In our planned experiment
\cite{Sondermann2007} we will use a ${}^{174}\mathrm{Yb}^{2+}$ ion as
a two-level system with ${}^1S_0$ and ${}^3P^0_1$ electronic levels as
the ground and the excited state respectively and no hyperfine
structure.  The atomic transition frequency $\omega_0\!\!=\!\!251.8$
nm is in the ultraviolet regime. The ion will be trapped at the focus
$f\!=\!2.1$mm of a metallic parabolic mirror, being one electrode of a
Paul trap. The rf needle-shaped electrode will come from the back of
the mirror through a small hole.  This trap design will ensure almost
full 4$\pi$ angle of ion-light interaction in the strong focusing
regime. The aberration corrections will be done using a diffractive
element located in front of the mirror. Since the ion has only one
decay channel and its dipole moment will be parallel to the mirror
axis we obtain free space geometry. There are several methods which
allow for single-photon pulse generation with the desired
spatio-temporal shape and spectral distribution. The first relies on
electro-optic modulation of a single-photon wave packet \cite{Harris}.
Another experimentally more accessible method applies a strongly
attenuated laser pulse containing $\overline{n}\!\!\ll \!\!1$ photons
on average. This technique is widely used in Quantum Key Distribution
\cite{QKD}.  We can shape a pulsed temporal mode electronically with
modulators starting from a continuous-wave laser.  Next we will turn
it into a radially polarized spatial doughnut mode.  After reflection
from the mirror surface its polarization, at the focal point will only
contribute to polarization parallel to the axis of the mirror and
therefore will excite a linear dipole oscillating parallel to this
axis.  Using this simpler method perfect coupling is achieved if the
probability of excitation matches the probability of finding a single
photon in the pulse.  In addition, as a third option one can generate
the properly shaped single-photon Fock state wave function
conditionally using photon pairs from parametric down conversion. This
method is similar to ghost imaging in the time domain.

Of course, none of these methods will produce an infinitely long
pulse. This is not an obstacle for our experiment however, because one
can truncate somewhat the exponential tail of the one-photon
wavepacket of Eq.(\ref{state1}).  For example, truncating the pulse to
a duration of five lifetimes the excitation probability can be as high
as 0.99. For quantum-storage applications it is straightforward to
expand this scheme to a lambda transition between two long lived
states \cite{Pinotsi}. Furthermore, efficient coupling in free space
opens the possibility for non-linear optics at the single photon
level.

\acknowledgements

M. S. acknowledges support by the A. von Humboldt foundation and by
MEN Grant No.~N202 021 32/0700. G. A. thanks T. Seligman and CIC in
Cuernavaca for their hospitality. The authors thank R. Maiwald,
U. Peschel, L. S\'anchez-Soto, S. Heugel, M. Sondermann and A. Villar
for discussions.

\end{document}